\documentclass[preprint,amsmath,amssymb,aps,superscriptaddress,nofootinbib,prd]{revtex4-1}
\usepackage{graphicx}
\usepackage{color}

\begin{document}


\title{The Muon and Tau Electric Dipole Moments in the B-L Supersymmetric Standard Model}

\author{Wen-Hui Zhang}\email{zwh$\_$0218@163.com}
\author{Jin-Lei Yang}\email{jlyang@hbu.edu.cn}
\affiliation{Department of Physics, Hebei University, Baoding, 071002, China}
\affiliation{Key Laboratory of High-precision Computation and Application of Quantum Field Theory of Hebei Province, Baoding, 071002, China}
\affiliation{Research Center for Computational Physics of Hebei Province, Baoding, 071002, China}

\author{Zhao-Feng Ge}
\affiliation{Radiation Monitoring Technical Center of Ministry of Ecology and Environment}
\affiliation{Key Laboratory of Radiation Environmental Monitoring, Ministry of Ecology and Environment}

\author{Yu-Li Yan}
\author{Yin-Jie Zhang}
\affiliation{Department of Physics, Hebei University, Baoding, 071002, China}
\affiliation{Key Laboratory of High-precision Computation and Application of Quantum Field Theory of Hebei Province, Baoding, 071002, China}
\affiliation{Research Center for Computational Physics of Hebei Province, Baoding, 071002, China}

\begin{abstract}

Recently proposed experiments are expected to significantly improve the measurement sensitivities of the electric dipole moments (EDMs) of muon ($d_\mu$) and tau ($d_\tau$). Given that theoretical predictions for $d_\mu$ and $d_\tau$ typically surpass those for the electron EDM, this work focuses on studying the contributions from the CP-violating (CPV) effects in the B-L supersymmetric (SUSY) standard model (B-LSSM) to $d_\mu$ and $d_\tau$. After considering the corrections from some two-loop diagrams, the contributions in the B-LSSM to the EDMs of charged leptons are presented analytically in general forms. The numerical results show that the traditional $\mu$-term in most SUSY models makes dominant contributions to $d_\mu$ and $d_\tau$, while the  B-LSSM specific CPV parameters also induce significant effects. It is found that across a substantial region of the B-LSSM parameter space, $d_\mu$ falls well within the projected sensitivity at Phase II of the proposed experiment, and $|d_\tau|$ can reach about $10^{-21}e\cdot\text{cm}$. 
\end{abstract}

\maketitle

\section{Introduction\label{sec1}}

The CP-violating (CPV) phases in Cabibbo-Kobayashi-Maskawa (CKM) matrix, which constitute the sole sources of CPV in the standard model (SM), have been observed precisely~\cite{Christenson:1964fg,Belle:2001zzw,BaBar:2002kla}. However, the contributions from the SM CPV interactions to the baryon asymmetry of the universe are too tiny to account for the experimental observations~\cite{Cooke:2013cba,Planck:2015fie}. Consequently, the introduction of new CPV sources in the new physics (NP) models is imperative. These new CPV sources may make significant contributions to the electric dipole moment (EDM) of a fundamental particle, which are highly suppressed by the small CKM phases in the SM~\cite{Pospelov:1991zt,Pospelov:2013sca,Yamaguchi:2020eub,Yamaguchi:2020dsy,Ema:2022yra}. This implies that the observations of non-zero fundamental particle EDMs would represent an unambiguous signal of NP~\cite{Engel:2013lsa,Chupp:2017rkp}, and corresponding theoretical studies may shed critical light on the origin of CPV.

Generally, most NP models predict the existence of new CPV sources and consequently large contributions to EDMs. For instance, the EDMs of electron, neutron and mercury have been extensively analyzed within the minimal supersymmetry (SUSY) model~\cite{Abel:2001vy,Ibrahim:2001ht,Cesarotti:2018huy}. The contributions in the minimal Left-Right model to the neutron and electron EDMs are studied in Ref.~\cite{Maiezza:2014ala,Bertolini:2019out}. Predictions for the electron EDM in the complex two-Higgs doublet model are presented in Ref.~\cite{Altmannshofer:2020shb}, where contributions from two-loop Barr-Zee diagrams~\cite{Barr:1990vd} are also included~\cite{Chun:2019oix}. In the Next-to-Minimal SUSY model, the prospects of measuring extra CPV sources through their contributions to EDMs are investigated~\cite{King:2015oxa}. Various two-loop contributions in the MSSM with $R-$parity violation are evaluated in Refs.~\cite{Chang:2000wf,Yamanaka:2012hm,Yamanaka:2012zq,Yamanaka:2012ep}. In the $U(1)_{B-L}$ extended MSSM (B-LSSM)~\cite{Khalil:2008ps,Elsayed:2011de,Elsayed:2012ec,Abdallah:2016vcn,Khalil:2015naa,DelleRose:2017uas,Yang:2020bmh,Yang:2021duj,Abdelalim:2020xfk,Khalil:2023jkm}, where $B,\;L$ represent baryon and lepton number respectively, we have previously explored the contributions from new CPV sources to the EDMs of neutron and heavy quarks~\cite{Yang:2019aao}, as well as to the electron EDM and electroweak baryogenesis~\cite{Yang:2020ebs}, the cancellations between different contributions of CPV phases are also analyzed.

Compared to the constraints on the EDMs of neutron, mercury and electron, the experimental constraints on the $\tau$ and $\mu$ lepton EDM $d_\tau,\;d_\mu$ are weaker due to the short lifetimes of them. The most stringent upper bound on $d_\tau$ is provided by the Belle II Collaboration as $-1.85\times10^{-17}e\cdot{\rm cm}<d_\tau<0.61\times10^{-17}e\cdot{\rm cm}$~\cite{Belle:2021ybo}, while the corresponding limit on $d_\mu$ is reported by the BNL muon $g-2$ experiment as $|d_\mu|<1.9\times10^{-19}e\cdot{\rm cm}$~\cite{Muong-2:2008ebm}. Both limits are much weaker than the constraint on the electron EDM, $|d_e|<4.1\times10^{-30}e\cdot{\rm cm}$~\cite{Roussy:2022cmp}. However, ongoing and projected experiments have the potential to improve the sensitivity to $d_\tau$, which can reach~\cite{Belle-II:2010dht,Belle-II:2018jsg,Aihara:2024zds,BESIII:2009fln,BESIII:2020nme,Bernreuther:2021uqm,CEPCStudyGroup:2018ghi,
CEPCPhysicsStudyGroup:2022uwl,Bodrov:2024wrw,Crivellin:2021spu}.
\begin{eqnarray}
&&d_\tau< 10^{-19}\;e\cdot{\rm cm}
\end{eqnarray}
Furthermore, utilizing the frozen-spin technique within a compact storage trap, a new experiment has been recently proposed to measure $d_\mu$, with expected sensitivities reaching~\cite{Crivellin:2018qmi,Adelmann:2025nev}:
\begin{eqnarray}
&&d_\mu<3\times10^{-21}\;e\cdot{\rm cm\;\;(Phase\;I)},\nonumber\\
&&d_\mu<6\times10^{-23}\;e\cdot{\rm cm\;\;(Phase\;II)}.
\end{eqnarray}
Consequently, there is strong motivation to extend theoretical investigations of EDMs to $d_\tau$ and $d_\mu$ within NP models, where these observables may be enhanced to levels accessible by next generation experiments. For instance, Ref.~\cite{Nakai:2025dmp} computes and analyzes predictions for $d_\tau$ within two benchmark models featuring different hypercharge assignments. Additionally, the contributions from light axion-like couplings of $\tau$ to $d_\tau$ are also analyzed~\cite{Huang:2025ghw}, they found the predicted $d_\tau$ approaches current experimental sensitivities and will be accessible to future measurements. Similarly, Ref.~\cite{Deka:2025qjc} utilizes the current upper bound and future projected sensitivity of $d_\mu$ to constrain dimension-six CP-violating operators within the SM effective field theory.

Based on our previous analyses of the neutron, heavy quarks, and electron EDMs within the B-LSSM, and motivated by the anticipated improvements in experimental sensitivities to $d_\mu$ and $d_\tau$, this work focuses on investigating the contributions from new CPV sources in the B-LSSM to these leptonic EDMs. Compared to the MSSM, the local gauge symmetry of the B-LSSM is extended by $U(1)_{B-L}$, introducing an additional $Z'$ gauge boson that can contribute to $d_\mu$ and $d_\tau$ at the two-loop level. Since the additional local gauge group is associated with the lepton number and the newly introduced scalar singlets acquire non-zero vacuum expectation values (VEVs), the observed neutrino oscillations can be naturally explained via the so-called Type-I seesaw mechanism~\cite{Minkowski:1977sc,Weinberg:1979sa}. In addition, the superpartners of these new scalar singlets can also serve as viable dark matter candidates~\cite{Yang:2023krd}. The MSSM predictions are now challenged by experiments, while the ones of B-LSSM can be relaxed significantly due to the additional structure. All these indicate that the B-LSSM is more attractive than the MSSM both theoretically and experimentally. Therefore,  this work comprehensively explores the constraints imposed by the current upper bounds on $d_\mu$ and $d_\tau$ upon the B-LSSM parameter space, and accesses whether the predicted  EDMs fall within the reach of future experimental measurements.

The paper is organized as follows: In Sec.~\ref{sec2}, the relevant mass matrices and interactions in the B-LSSM are introduced, and the contributions to charged lepton EDMs are calculated analytically. The numerical results are presented and analyzed in Sec.~\ref{sec3}. A brief summary is given in Sec.~\ref{sec4}. The lengthy formulae are collected in the Appendix.

\section{The B-LSSM and calculations of $d_l$\label{sec2}}

In this section, we outline the relevant framework of the B-LSSM and present the analytical calculations for the charged lepton EDMs, $d_l$.

\subsection{The relevant mass matrices and interactions}

The scalar sector of the B-LSSM is extended by two scalar singlets
\begin{eqnarray}
&&\eta\sim(1,1,0,-1),\;\;\bar\eta\sim(1,1,0,1),
\end{eqnarray}
where the charges in the brackets denote the quantum numbers under $SU(3)_C,$ $ SU(2)_L,$ $ U(1)_Y,$ $U(1)_{B-L}$, respectively. The additional $U(1)_{B-L}$ local gauge symmetry is spontaneously broken when the two scalar singlet fields acquire non-zero VEVs:
\begin{eqnarray}
&&\eta=\frac{1}{\sqrt2}(u_\eta+\phi_\eta+i\sigma_\eta),
\qquad\;\quad\;\bar\eta=\frac{1}{\sqrt2}(u_{\bar\eta}+\phi_{\bar\eta}+i\sigma_{\bar\eta}).
\end{eqnarray}
where $\tan\beta'\equiv u_{\bar\eta}/u_\eta$ and $u^2\equiv u_\eta^2+u_{\bar\eta}^2$. The spontaneous breaking of $U(1)_{B-L}$ generates Majorana masses for the right-handed neutrinos. By combining these with the Dirac masses arising from the Higgs doublet VEVs, the Type-I seesaw mechanism is naturally realized.

At the one-loop level, the dominant contributions to $d_\mu$ and $d_\tau$ come from diagrams mediated by sleptons-neutralinos and charginos-sneutrinos in the loop. The slepton mass matrix can be written as
\begin{eqnarray}
&&m_{\tilde l}^2=
\left(\begin{array}{cc}m_{\tilde l_L}^2,&\frac{1}{\sqrt2}Y_e^\dagger(v_1 A_l^\dagger- v_2 \mu)\\ \frac{1}{\sqrt2}Y_e (v_1 A_l-v_2 \mu^*),&m_{\tilde l_R}^2\end{array}\right),\label{eq-SL-mass}
\end{eqnarray}
where
\begin{eqnarray}
&&m_{\tilde l_L}^2=\frac{1}{8}\Big[(g_1^2+g_{YB}^2-g_2^2+g_B g_{YB})(v_1^2-v_2^2)+2g_B(g_B+g_{YB})(u_\eta^2-u_{\bar \eta}^2)\Big]\nonumber\\
&&\qquad\quad+m_{\tilde l}^2+\frac{1}{2}v_1^2Y_eY_e^\dagger,\nonumber\\
&&m_{\tilde l_R}^2=\frac{1}{8}\Big[(2g_1^2+2g_{YB}^2+g_B g_{YB})(v_2^2-v_1^2)+2g_B(g_B+2g_{YB})(u_{\bar \eta}^2-u_\eta^2)\Big]\nonumber\\
&&\qquad\quad+m_{\tilde e}^2+\frac{1}{2}v_1^2Y_eY_e^\dagger.
\end{eqnarray}
Similar to the MSSM, $g_1$ and $g_2$ are the standard gauge coupling constants, $v_1, v_2$ are the VEVs of the Higgs doublets, $A_l$ denotes the trilinear coupling constant, $\mu$ is the $\mu$ term coupling, $m_{\tilde l},\;m_{\tilde e}$ are the soft SUSY-breaking mass terms.  Additionally, $Y_e$ is the Yukawa coupling matrix of charged leptons. Furthermore, $g_B$ is the gauge coupling constant associated with $U(1)_{B-L}$, and $g_{YB}$ is the gauge coupling arising from the gauge kinetic mixing~\cite{Holdom:1985ag,Matsuoka:1986ig,delAguila:1987st,delAguila:1988jz,Foot:1991kb,Babu:1997st}. The squared mass matrices for CP-odd and CP-even sneutrinos can be written as
\begin{eqnarray}
&&m_{\tilde \nu_{\rm odd},\tilde \nu_{\rm even}}^2=
\left(\begin{array}{cc}m_{\tilde \nu_L}^2,&\frac{1}{\sqrt2}v_2 \mathcal{R}(A_\nu)\\ \frac{1}{\sqrt2}v_2 \mathcal{R}(A_\nu),&m_{\tilde \nu_{{\rm odd,even} R}}^2\end{array}\right),
\end{eqnarray}
where
\begin{eqnarray}
&&m_{\tilde \nu_L}^2=\frac{1}{8}\Big[(g_1^2+g_{YB}^2+g_2^2+g_B g_{YB})(v_1^2-v_2^2)+2g_B(g_B+g_{YB})(u_\eta^2-u_{\bar \eta}^2)\Big]+m_{\tilde l}^2,\nonumber\\
&&m_{\tilde \nu_{{\rm odd,even}R}}^2=\frac{1}{8}\Big[g_B g_{YB}(v_2^2-v_1^2)+2g_B^2(u_{\bar \eta}^2-u_\eta^2)\Big]-\sqrt2\Big[u_{\eta}\mathcal{R}(A_R)-\sqrt2u_{\eta}^2 Y_R^\dagger Y_R\Big]\nonumber\\
&&\qquad\qquad\quad\;\;+m_{\tilde \nu}^2\pm\sqrt2 u_{\bar \eta}\mathcal{R}(Y_R\mu_\eta^*).
\end{eqnarray}
Here, the $\pm$ signs correspond to CP-odd and CP-even sneutrinos respectively. $\mathcal{R}$ denotes the real part. $Y_R$ is the Majorana coupling matrix of right-handed neutrinos while the Dirac coupling matrix is taken to be zero for simplicity. $A_\nu,\;A_R$ denote the relevant trilinear coupling constants. It is worth noting that all matrix elements of the sneutrinos are real; therefore, they do not significantly affect the numerical results for $d_\mu$ and $d_\tau$. To simplify the parameter space, we take $Y_R={\rm diag}(0.1,0.1,0.1),\;A_\nu=A_R={\rm diag}(0.1,0.1,0.1)\;{\rm TeV}$ in the following computations.

The mass matrix of charginos in the B-LSSM is same as the one in the MSSM, while the mass matrix of neutralinos on the basis $(\tilde{\lambda}_{B},\tilde{\lambda}_{W},\tilde{\lambda}_{H^0_d},\tilde{\lambda}_{H^0_u},\tilde{\lambda}_{B'},\tilde{\lambda}_{\eta},\tilde{\lambda}_{\bar{\eta}})$ can be written as
\begin{eqnarray}
m_{\tilde{\chi}^0}=\left(\begin{array}{ccccccc}
M_1 & 0 & -\frac{1}{2}g_1 v_1 & \frac{1}{2}g_1v_2 & M_{BB'} & 0 & 0 \\ 0 & M_2 & \frac{1}{2}g_2 v_1 & -\frac{1}{2}g_2v_2 & 0 & 0 & 0 \\ -\frac{1}{2}g_1 v_1 & \frac{1}{2}g_2 v_1 & 0 & -\mu & -\frac{1}{2}g_{YB} v_1 & 0 & 0 \\ \frac{1}{2}g_1 v_2 & -\frac{1}{2}g_2 v_2 & -\mu & 0 & \frac{1}{2}g_{YB} v_2 & 0 & 0 \\ M_{BB'} & 0 & -\frac{1}{2}g_{YB} v_1 & \frac{1}{2}g_{YB} v_2 & M_{BL} & -g_Bu_\eta & g_Bu_{\bar{\eta}}\\ 0 & 0 & 0 & 0 & -g_Bu_\eta & 0 & -\mu_\eta \\ 0 & 0 & 0 & 0 & g_Bu_{\bar{\eta}} & -\mu_\eta & 0 \\
\end{array}
\right).
\end{eqnarray}
Obviously, there are three additional neutralinos in the B-LSSM, which can make contributions to $d_\mu$ and $d_\tau$ at the one-loop level. Correspondingly, there are also three new CPV sources $M_{BB'},\;M_{BL},\;\mu_\eta$ in the model, they can affect the theoretical predictions on $d_\mu,\;d_\tau$ through the one-loop level contributions.

The new scalars and $Z'$ gauge boson can also contribute to the charged lepton EDMs through the Barr-Zee diagrams. The squared Higgs mass matrix after considering the two-loop effective potential corrections from top and stop quarks can be found in our previous work~\cite{Yang:2024fol}, which is used to carry out the numerical computations. The squared mass of new gauge boson $Z'$ can be estimated by
\begin{eqnarray}
M_{Z'}\approx \frac{1}{2}g_B u,
\end{eqnarray}
where the terms proportional to $v$ are neglected. All involved coupling constants are collected in the Appendix.

\subsection{The contributions to $d_l$ in the model}

\begin{figure}
\setlength{\unitlength}{1mm}
\centering
\includegraphics[width=6in]{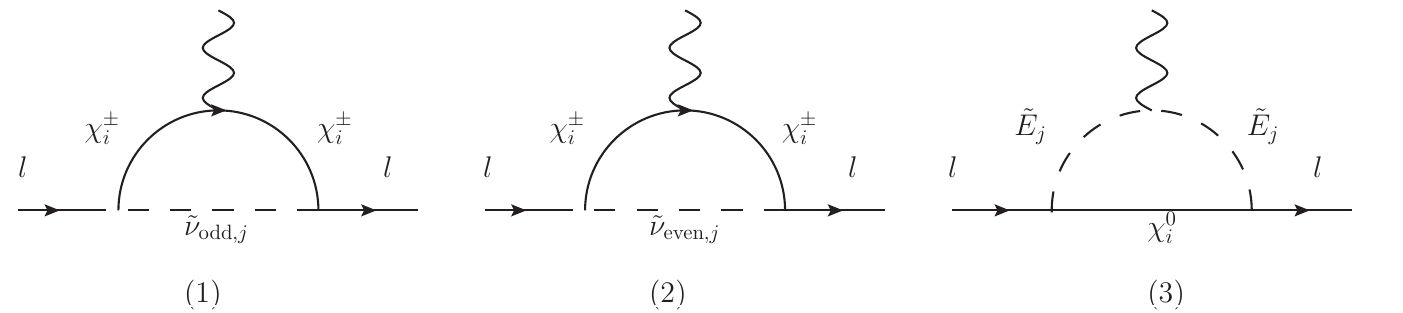}
\vspace{0cm}
\caption[]{The one-loop level diagrams contributing to the charged lepton EDMs.}
\label{Fig1}
\end{figure}

The effective Lagrangian for the charged lepton EDMs is described by the following operator:
\begin{eqnarray}
&&\mathcal{L}_{\rm eff}=-\frac{d_l}{2}(\bar l\sigma^{\mu\nu}i\gamma_5 l)F_{\mu\nu}
\end{eqnarray}
where $l=e,\mu,\tau$ denote the charged lepton fields, $\sigma^{\mu\nu}=i[\gamma^\mu,\gamma^\nu]/2$ with $\gamma^\mu$ being the Dirac matrix, and $F_{\mu\nu}$ is the electromagnetic field strength tensor.  The leading order contributions to $d_l$ in the B-LSSM arise from the one-loop diagrams, which are plotted in Fig.~\ref{Fig1}. The resulting exoressions can be written as~\cite{Cheung:2009fc}
\begin{eqnarray}
&&d_l^{(1)}=\sum_{i,j} \frac{-eM_{\chi^\pm_i}}{16\pi^2 M_{\nu_{{\rm odd},j}}^2}\mathcal{I}\Big(C_{\bar l\chi^\pm_i\nu_{{\rm odd},j}}^LC_{\bar l\chi^\pm_i\nu_{{\rm odd},j}}^R\Big)f_1\Big(\frac{M_{\chi^\pm_i}^2}{M_{\nu_{{\rm odd},j}}^2}\Big),\\
&&d_l^{(2)}=\sum_{i,j} \frac{-eM_{\chi^\pm_i}}{16\pi^2 M_{\nu_{{\rm even},j}}^2}\mathcal{I}\Big(C_{\bar l\chi^\pm_i\nu_{{\rm even},j}}^LC_{\bar l\chi^\pm_i\nu_{{\rm even},j}}^R\Big)f_1\Big(\frac{M_{\chi^\pm_i}^2}{M_{\nu_{{\rm even},j}}^2}\Big),\\
&&d_l^{(3)}=\sum_{i,j} \frac{-eM_{\chi^0_i}}{16\pi^2 M_{\tilde E_j}^2}\mathcal{I}\Big(C_{\bar l\chi^0_i \tilde E_j}^LC_{\bar l\chi^0_i \tilde E_j}^R\Big)f_2\Big(\frac{M_{\chi^0_i}^2}{M_{\tilde E_j}^2}\Big),
\end{eqnarray}
where $\mathcal{I}$ denotes the imaginary part, $M_{\nu_{{\rm odd}}},\;M_{\nu_{{\rm even}}},\;M_{\chi^\pm},\;M_{\chi^0},\;M_{\tilde E}$ denote the physical masses of the CP-odd sneutrinos, CP-even sneutrinos, charginos, neutralinos and sleptons, respectively. Furthermore, $C_{XYZ}^{L(R)}$ denotes the left (right)-handed coupling constant of the interactions between $XYZ$ which can be found in the Appendix, and
\begin{eqnarray}
&&f_1(x)=\frac{1}{2(1-x)^2}[3-x+\frac{2\ln\;x}{1-x}],\\
&&f_2(x)=\frac{1}{2(1-x)^2}[1+x+\frac{2x\ln\;x}{1-x}].
\end{eqnarray}

\begin{figure}
\setlength{\unitlength}{1mm}
\centering
\includegraphics[width=5in]{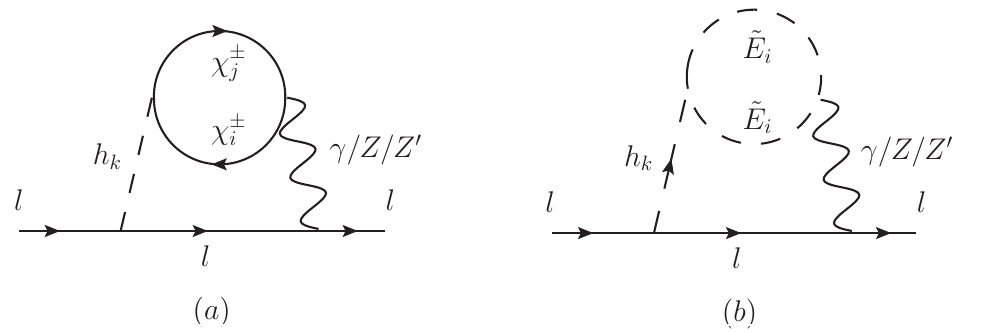}
\vspace{0cm}
\caption[]{The considered two-loop level diagrams contributing to the charged lepton EDMs, where photon is attached by all possible ways.}
\label{Fig2}
\end{figure}

The two-loop level contributions to $d_l$ considered in this work are plotted in Fig.~\ref{Fig2}, where the external photon is attached in all possible ways. Clearly, the new scalars and $Z'$ gauge boson can also make contributions, and the resulting $d_l$ can be correspondingly written as ~\cite{Yamanaka:2012qn,Altmannshofer:2025jkk}
\begin{eqnarray}
&&d_l^{(a)-\gamma}=\frac{-e^2m_l}{16\pi^4}\sum_{i,j,k}\mathcal{I}\Big(C_{h_k\bar\chi^\pm_i\chi^\pm_j}C_{h_k\bar ll}^*\Big)\Big[\ln \Big(\frac{M_{h_k}^2}{m_l^2}\Big)-2+\frac{m_l^2}{M_{h_k}^2}\Phi\Big(\frac{m_l^2}{M_{h_k}^2}\Big)\Big]\frac{1}{M_{h_k}^2},
\end{eqnarray}

\begin{eqnarray}
&&d_l^{(a)-Z}=\frac{-e^2m_l}{64\pi^4}\sum_{i,j,k}\mathcal{I}\Big[C_{h_k\bar\chi^\pm_i\chi^\pm_j}C_{h_k\bar ll}^*(C_{Z\bar ll}^L+C_{Z\bar ll}^R)(C_{Z\bar\chi^\pm_i\chi^\pm_j}^L+C_{Z\bar\chi^\pm_i\chi^\pm_j}^R)\Big]\frac{1}{M_{h_k}^2}\nonumber\\
&&\qquad\qquad\times\Big[\ln\Big(\frac{M_{h_k}^2}{M_Z^2}\Big)-\frac{m_l^2}{M_Z^2}\Phi\Big(\frac{m_l^2}{M_Z^2}\Big)+\frac{m_l^2}{M_{h_k}^2}\Phi\Big(\frac{m_l^2}{M_{h_k}^2}\Big)\Big],\\
&&d_l^{(b)-\gamma}=\frac{-e^2}{128\pi^4}\sum_{i,k}\mathcal{I}\Big(C_{h_k\tilde E_i\tilde E_i}C_{h_k\bar ll}^*\Big)\Big[\ln\Big(\frac{M_{h_k}^2}{M_{\tilde E_i}^2}\Big)+\frac{M_{\tilde E_i}^2}{M_{h_k}^2}\Phi\Big(\frac{M_{\tilde E_i}^2}{M_{h_k}^2}\Big)\Big]\frac{1}{M_{h_k}^2},\\
&&d_l^{(b)-Z}=\frac{-e}{128\pi^4}\sum_{i,k}\mathcal{I}\Big[C_{h_k\tilde E_i\tilde E_i}C_{h_k\bar ll}^*C_{Z\tilde E_i\tilde E_i}(C_{Z\bar ll}^L+C_{Z\bar ll}^R)\Big]\frac{1}{M_{h_k}^2-M_Z^2}\nonumber\\
&&\qquad\qquad\times \Big[\ln \Big(\frac{M_{h_k}^2}{M_Z^2}\Big)-\frac{M_{\tilde E_i}^2}{M_Z^2}\Phi\Big(\frac{M_{\tilde E_i}^2}{M_Z^2}\Big)+\frac{M_{\tilde E_i}^2}{M_{h_k}^2}\Phi\Big(\frac{M_{\tilde E_i}^2}{M_{h_k}^2}\Big)\Big],
\end{eqnarray}
where $d_l^{(a)-Z'}, d_l^{(b)-Z'}$ can be obtained by replacing $Z$ boson with a $Z'$ boson in the expressions for $d_l^{(a)-Z}, d_l^{(b)-Z}$. Here, $m_l$ denotes the mass of charged lepton $l$, and
\begin{eqnarray}
&&\Phi(x)=\frac{1}{\sqrt{1-4x}}\Big[\frac{\pi^2}{3}-\ln^2(x)+2\ln^2\Big(\frac{1-\sqrt{1-4x}}{2}\Big)-4{\rm Li}_2\Big(\frac{1-\sqrt{1-4x}}{2}\Big)\Big].
\end{eqnarray}

Actually, the CPV parameters in the squark sector can also make contributions to $d_\mu,\;d_\tau$ through the two-loop diagrams in Fig.~\ref{Fig2}, but they will make significant contributions to the neutron, mercury and heavy quark EDMs at the one-loop level, which are highly suppressed by the corresponding experimental upper bounds. Hence, We can neglect safely the contributions from the CPV parameters in the squark sector to $d_\mu,\;d_\tau$ at the two-loop level. Finally, the EDM of charged lepton predicted dominantly in the B-LSSM can be calculated by
\begin{eqnarray}
&&d_l=d_l^{(1)}+d_l^{(2)}+d_l^{(3)}+d_l^{(a)-\gamma}+d_l^{(b)-\gamma}+d_l^{(a)-Z}+d_l^{(b)-Z}+d_l^{(a)-Z'}+d_l^{(b)-Z'}.
\end{eqnarray}

\section{Numerical results of $d_\mu,\;d_\tau$ in the B-LSSM\label{sec3}}

In this section, we present the numerical predictions for $d_\mu$ and $d_\tau$ in the B-LSSM. For the relevant SM input parameters, we take $m_W=80.385\;{\rm GeV},\;m_Z=90.1876\;{\rm GeV},\;\alpha_{em}(m_Z)=1/128.9,\;\alpha_s(m_Z)=0.118$. The measured SM-like Higgs mass is taken as ~\cite{ParticleDataGroup:2024cfk}
\begin{eqnarray}
&&m_h=125.09\pm0.24{\rm GeV}.\label{hma}
\end{eqnarray}
The newly introduced $Z'$ gauge boson,  associated with the $U(1)_{B-L}$ gauge group, can also contribute via the two-loop diagrams. The ratio between the $Z'$ mass, $M_{Z'}\approx\frac{1}{2}g_B u$, and its gauge coupling is subject to a strict lower bound at the $99\%$ confidence level (CL) as~\cite{Carena:2004xs,Cacciapaglia:2006pk,ATLAS:2016cyf}:
\begin{eqnarray}
&&M_{Z'}/g_{_B}\geq6\;{\rm TeV}\;.\label{eq27}
\end{eqnarray}

For the other relevant parameters in the B-LSSM, we take the following benchmark points for simplicity to carry out the numerical computations
\begin{eqnarray}
&&\tan\beta=10,\;\tan\beta'=1.15,\;g_B=0.4,\;g_{YB}=-0.4,\;M_{Z'}=5\;\text{TeV},\nonumber\\
&&m_{\tilde l}=m_{\tilde e}=m_{\tilde\nu}=\text{diag}(M_{\tilde L},M_{\tilde L},\;M_{\tilde L}),\;M_{\tilde L}=0.5\;\text{TeV},\;M_{\tilde t}=M_{\tilde b}=2\;\text{TeV},\nonumber\\
&&M_1=M_2=\mu=0.6e^{i\theta_\mu}\;\text{TeV},\;A_\mu=0.1e^{i\theta_{A_\mu}}\;\text{TeV},\;A_\tau=0.1e^{i\theta_{A_\tau}}\;\text{TeV},\nonumber\\
&&M_{BB'}=0.6e^{i\theta_{BB'}}\;\text{TeV},\;M_{BL}=0.6e^{i\theta_{BL}}\;\text{TeV},\;\mu_\eta=0.6e^{i\theta_{\mu_\eta}}\;\text{TeV},\nonumber\\
&&\theta_\mu=0,\;\theta_{A_\mu}=0,\;\theta_{BB'}=0,\;\theta_{BL}=0,\;\theta_{\mu_\eta}=0.\label{BMP}
\end{eqnarray}
The parameters in the first three lines also appeared in most SUSY models, while the ones in the penultimate line are the B-LSSM specific parameters. $M_{\tilde t},\;M_{\tilde b}$ denote the stop and sbottom masses, respectively, which are used to calculate two-loop corrections to the SM-like Higgs boson mass. Unless we specify the parameter's value or the parameter as variable, Eq.~(\ref{BMP}) is taken as input in the numerical computations. The parameters fixed in Eq.~(\ref{BMP}) can well satisfy the constrains from LHC experimental data~\cite{Basso:2015xna}, high-precision measured $\text{Br}(\bar B\rightarrow X_s\gamma)$, $\text{Br}(B_s^0\rightarrow \mu^+\mu^-)$~\cite{Yang:2018fvw}, the direct searches of squarks, and the observed Higgs signal~\cite{Un:2016hji}. All flavor violating parameters are assumed to be zero without losing generality.

\subsection{The effects of parameters in general SUSY models}

\begin{figure}
\setlength{\unitlength}{1mm}
\centering
\includegraphics[width=3.1in]{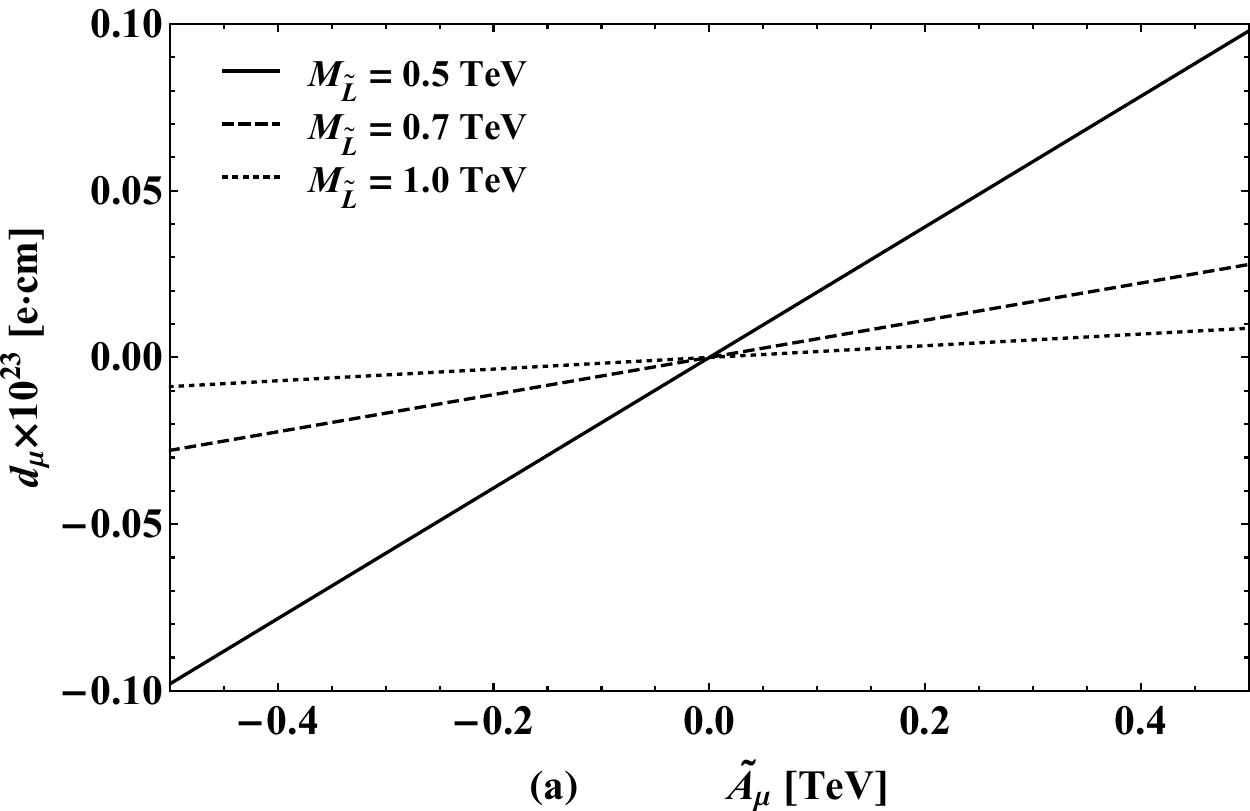}%
\vspace{0.5cm}
\includegraphics[width=3.1in]{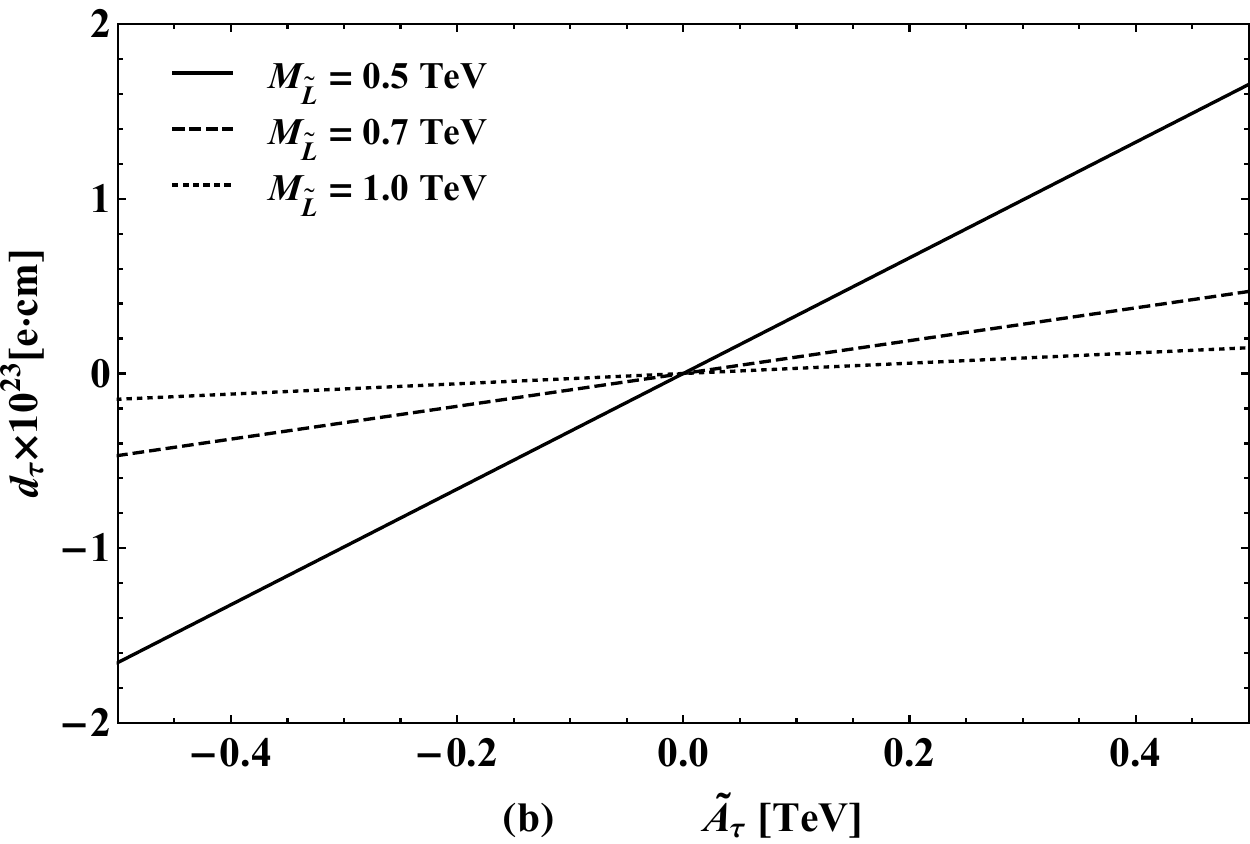}
\vspace{0cm}
\par
\hspace{-0.in}
\includegraphics[width=3.1in]{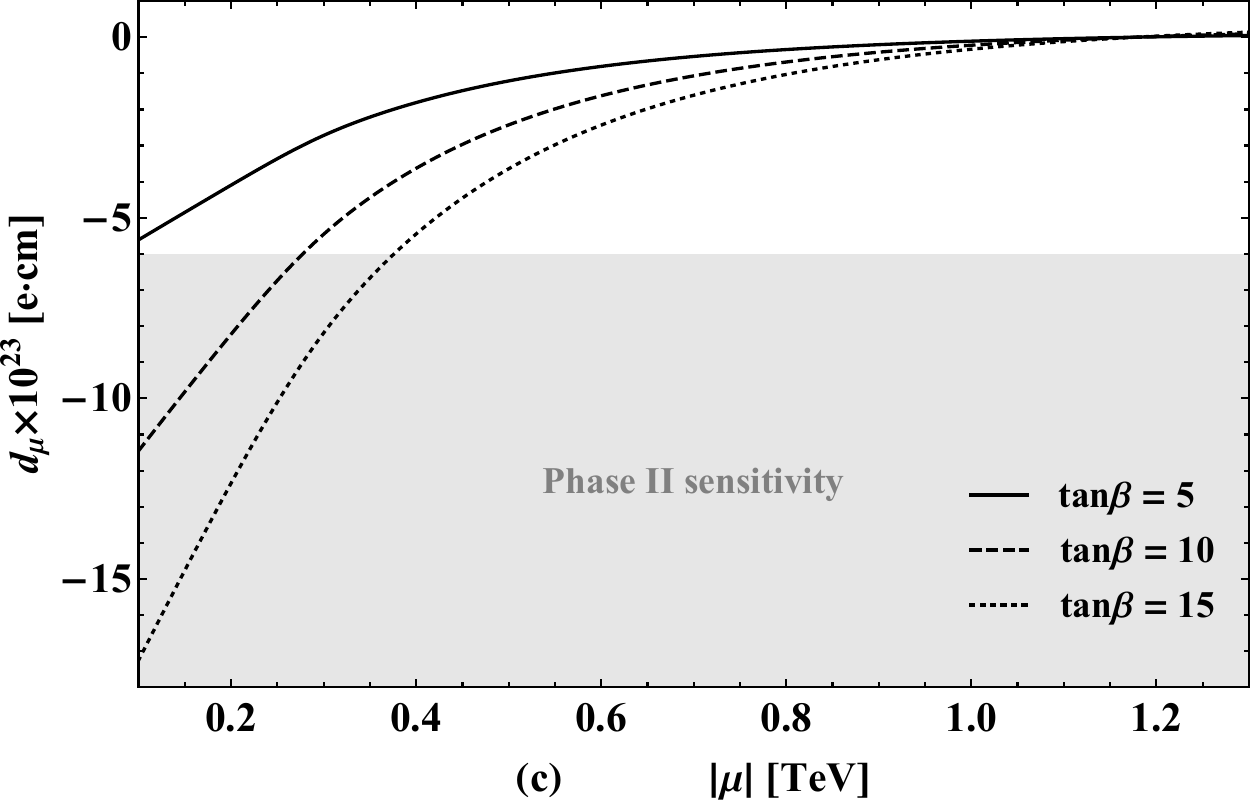}%
\vspace{0.5cm}
\includegraphics[width=3.1in]{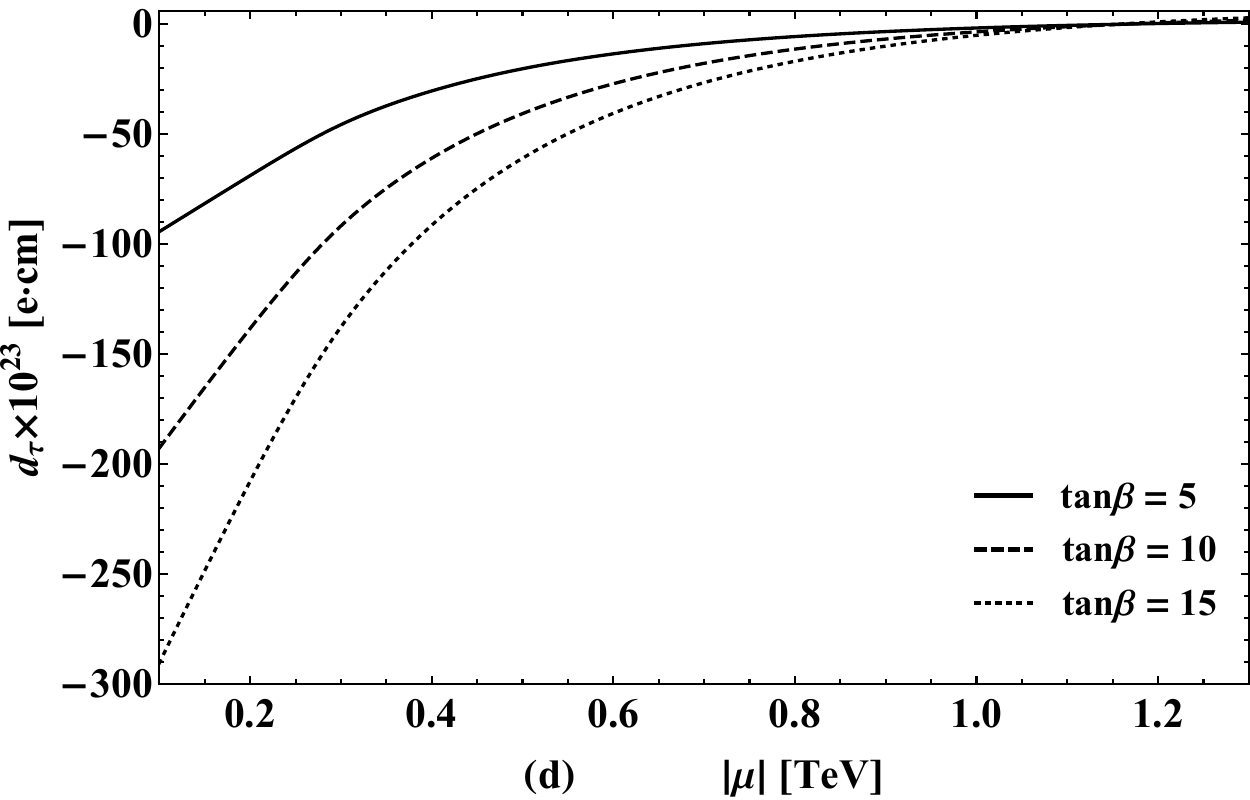}
\vspace{0cm}
\caption[]{The numerical results of $d_\mu$ versus $\tilde A_\mu$ (a) with $\theta_{A_\mu}=0.5\pi$, $d_\tau$ versus $\tilde A_\tau$ (b) with $\theta_{A_\tau}=0.5\pi$ and $d_\mu$ (c), $d_\tau$ (d) versus $|\mu|$ with $\theta_{\mu}=0.5\pi$, the shaded area denotes the sensitivity can be reached in phase II.}
\label{Fig3}
\end{figure}

Since the CPV parameters $\mu$ and $A_l\;(l=e,\mu,\tau)$ are ubiquitous in most SUSY models, we first explore their effects on $d_\mu$ and $d_\tau$. In Figs.~\ref{Fig3}(a) and ~\ref{Fig3}(b), we set the CPV phases of the trilinear scalar couplings to their maximal values, $\theta_{A_\mu} = \theta_{A_\tau} = 0.5\pi$. These figures illustrate the evolution of $d_\mu$ and $d_\tau$ as functions of the real parameters $\tilde A_\mu$ and $\tilde A_\tau$, respectively, where $A_l\equiv \tilde A_l e^{i\theta_{A_l}}$. The solid, dashed, and dotted lines correspond to slepton mass parameters of $M_{\tilde L}=0.5,\;0.7,\;1.0\;\text{TeV}$, respectively.

Evidently, both $d_\mu$ and $d_\tau$ exhibit a clear linear dependence on their respective trilinear parameters $\tilde A_l$, vanishing exactly at $\tilde A_l=0$. This behavior indicates that $A_l$ serves as a primary direct CPV source for these loop-induced EDMs. Furthermore, as the slepton mass parameter $M_{\tilde{L}}$ increases, the slopes of these linear dependencies are strongly suppressed, clearly reflecting the decoupling effect of SUSY particles. A comparison between Figs.~\ref{Fig3} (a) and (b) further reveals that the magnitude of $d_\tau$ is much larger than that of $d_\mu$, because the $A_l$-induced left-right mixing term in the slepton mass matrix is proportional to the corresponding Yukawa couplings.

Figs.~\ref{Fig3} (c) and (d) further investigate the CPV effects induced by the Higgsino mass parameter $\mu$, whose CPV phase serves as a crucial source for generating the baryon asymmetry of the universe via electroweak baryogenesis. In these figures, we fix the CPV phase at $\theta_\mu = 0.5\pi$ and illustrate the dependence of the lepton EDMs on $|\mu|$ for different values of $\tan\beta$. The solid, dashed, and dotted lines denote the results of $\tan\beta = 5, 10, 15$, respectively. It can be seen that as $|\mu|$ increases, leading to heavier Higgsinos, the magnitudes of $d_\mu$ and $d_\tau$ asymptotically approach zero, which clearly confirms the decoupling behavior of heavy SUSY particles. On the other hand, the absolute magnitudes of the EDMs are significantly enhanced by larger values of $\tan \beta$. This is because a larger $\tan\beta$ substantially increases the Yukawa couplings of the down-type fermions, thereby greatly amplifying the theoretical predictions for the EDMs. Notably, the gray shaded area in Figs.~\ref{Fig3} (c) indicates the anticipated sensitivity of the future Phase II experiment for $d_\mu$. The numerical results demonstrate that in the parameter space characterized by a large $\tan\beta$ (e.g., $\tan\beta>10$) and relatively light Higgsinos ($|\mu| \lesssim 0.4\text{ TeV}$), the theoretical prediction for $d_\mu$ falls entirely within the reach of Phase II.

\subsection{The effects of B-LSSM specific parameters}

\begin{figure}
\setlength{\unitlength}{1mm}
\centering
\includegraphics[width=3.1in]{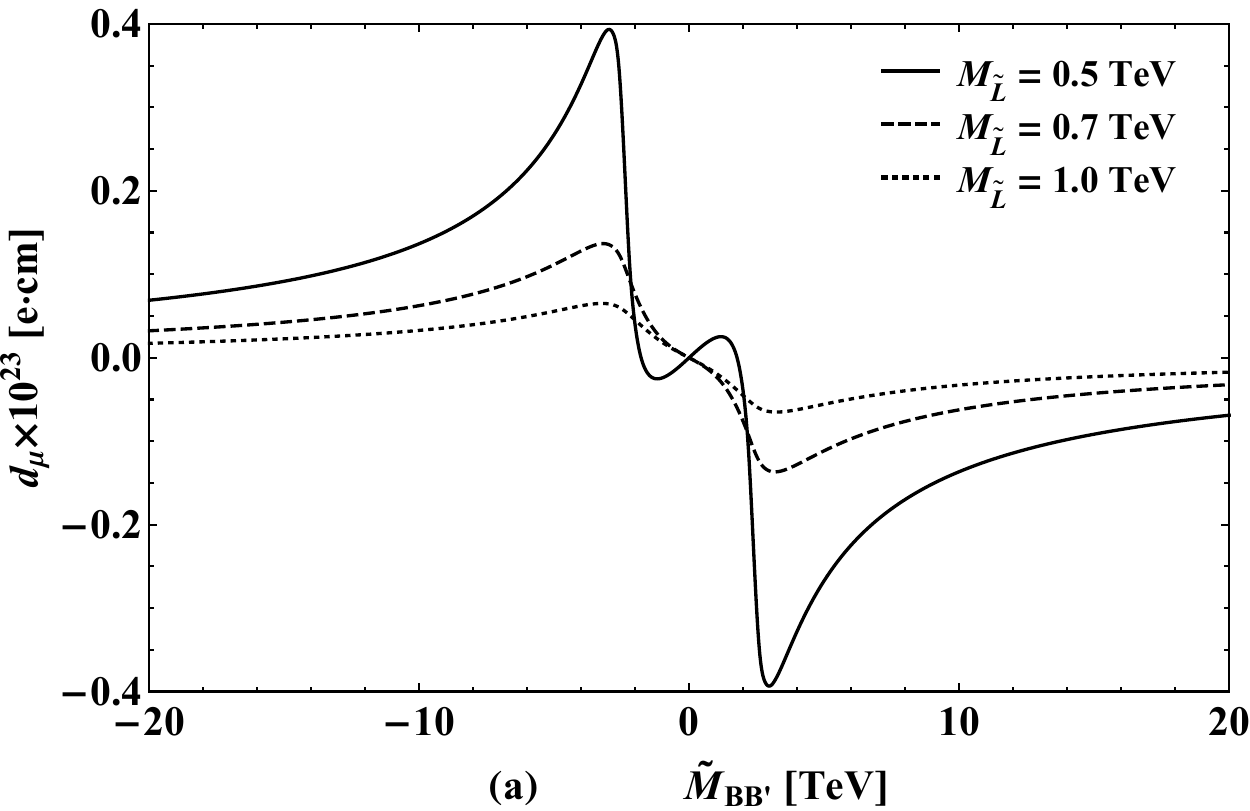}%
\vspace{0.5cm}
\includegraphics[width=3.1in]{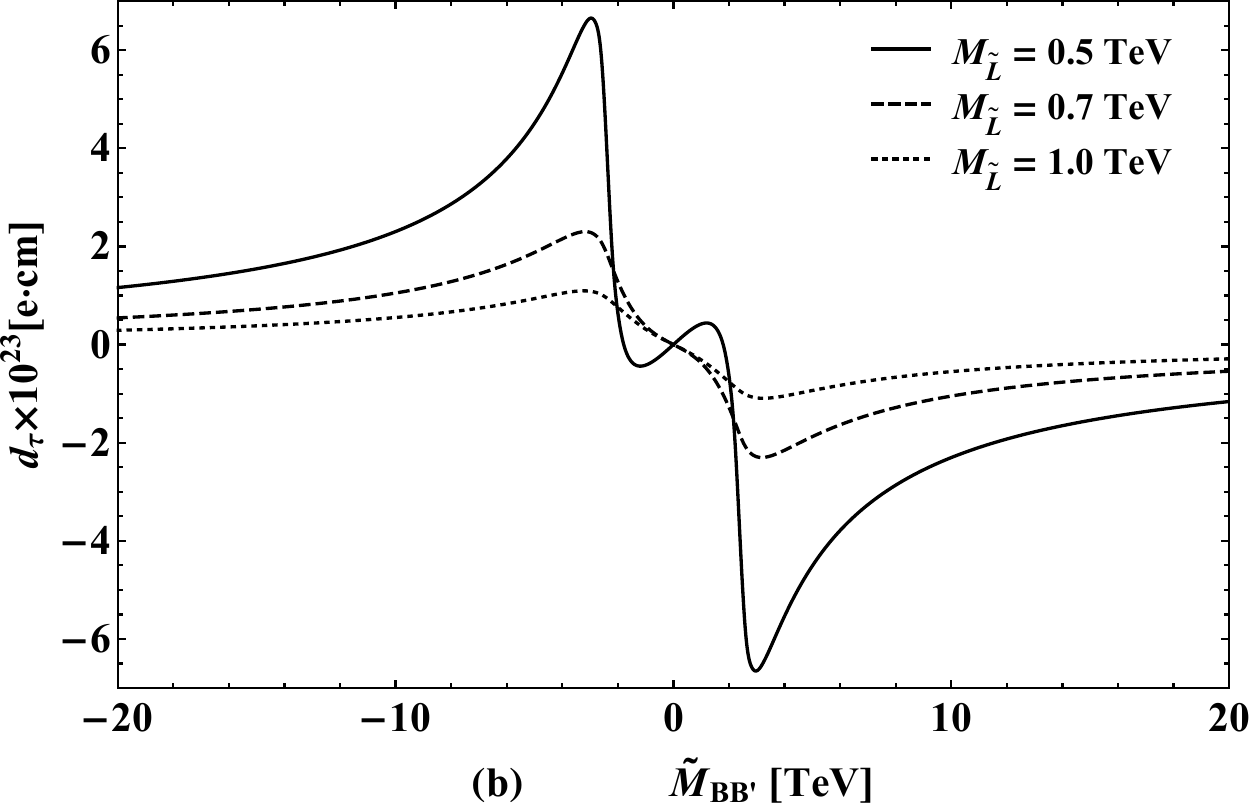}
\vspace{0cm}
\par
\hspace{-0.in}
\includegraphics[width=3.1in]{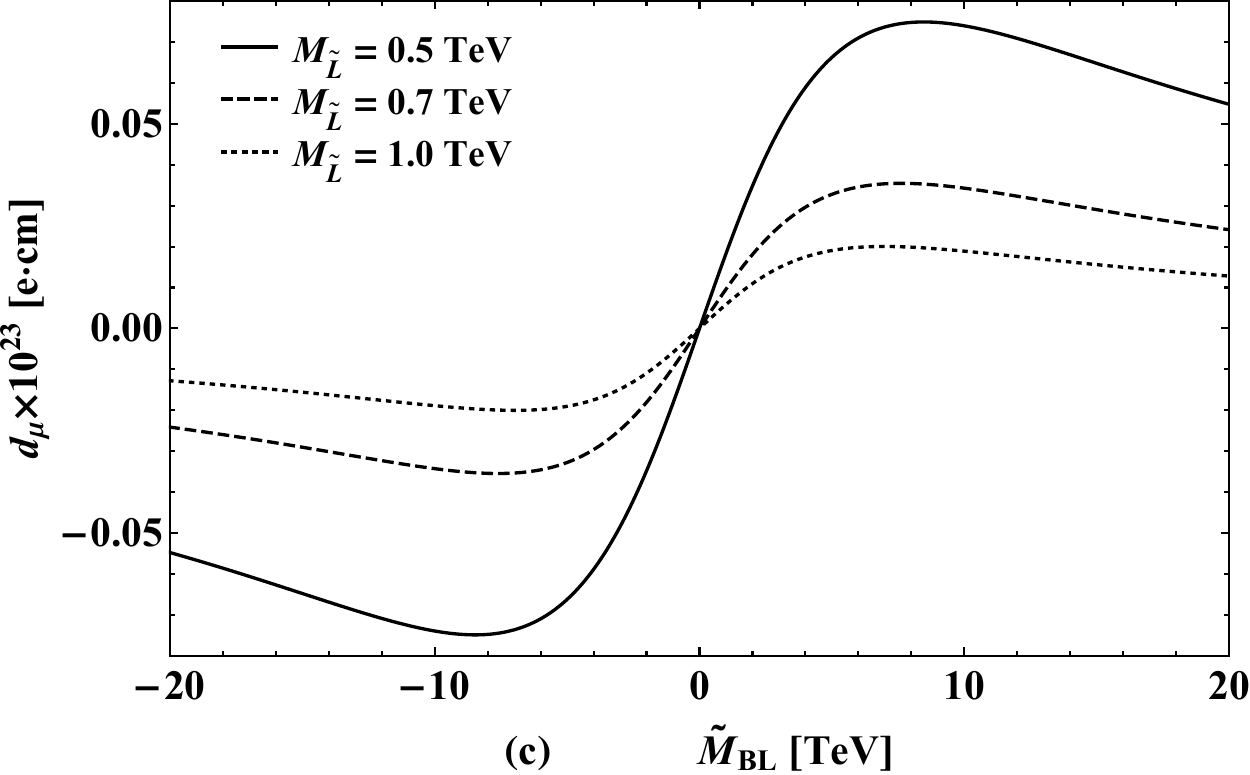}%
\vspace{0.5cm}
\includegraphics[width=3.1in]{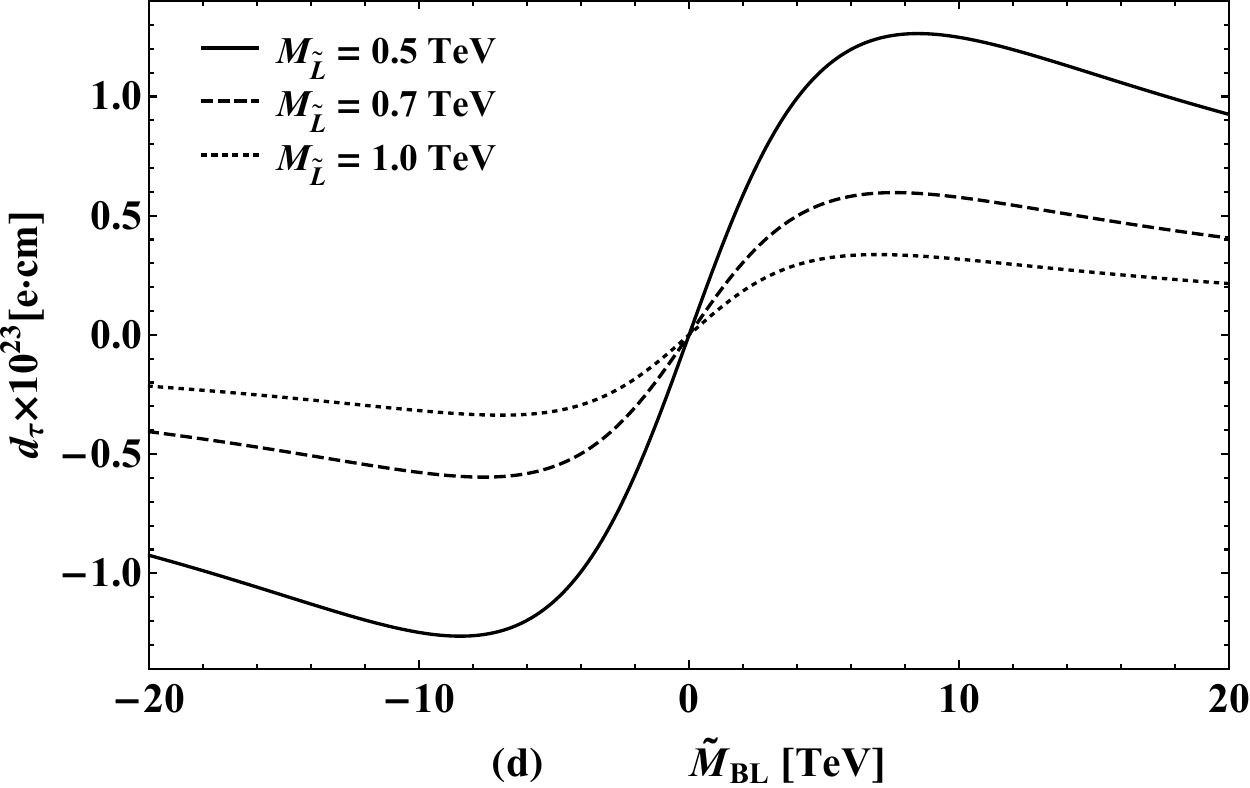}
\vspace{0cm}
\par
\hspace{-0.in}
\includegraphics[width=3.1in]{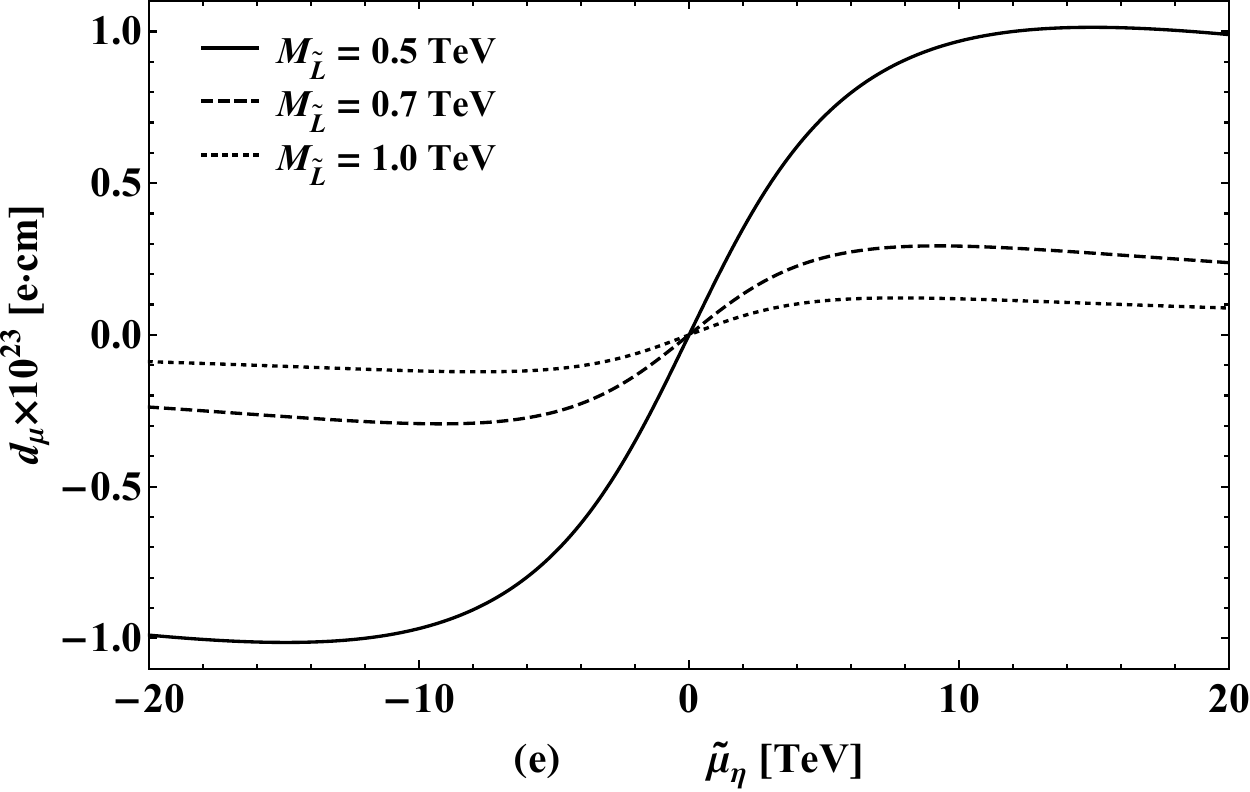}%
\vspace{0.5cm}
\includegraphics[width=3.1in]{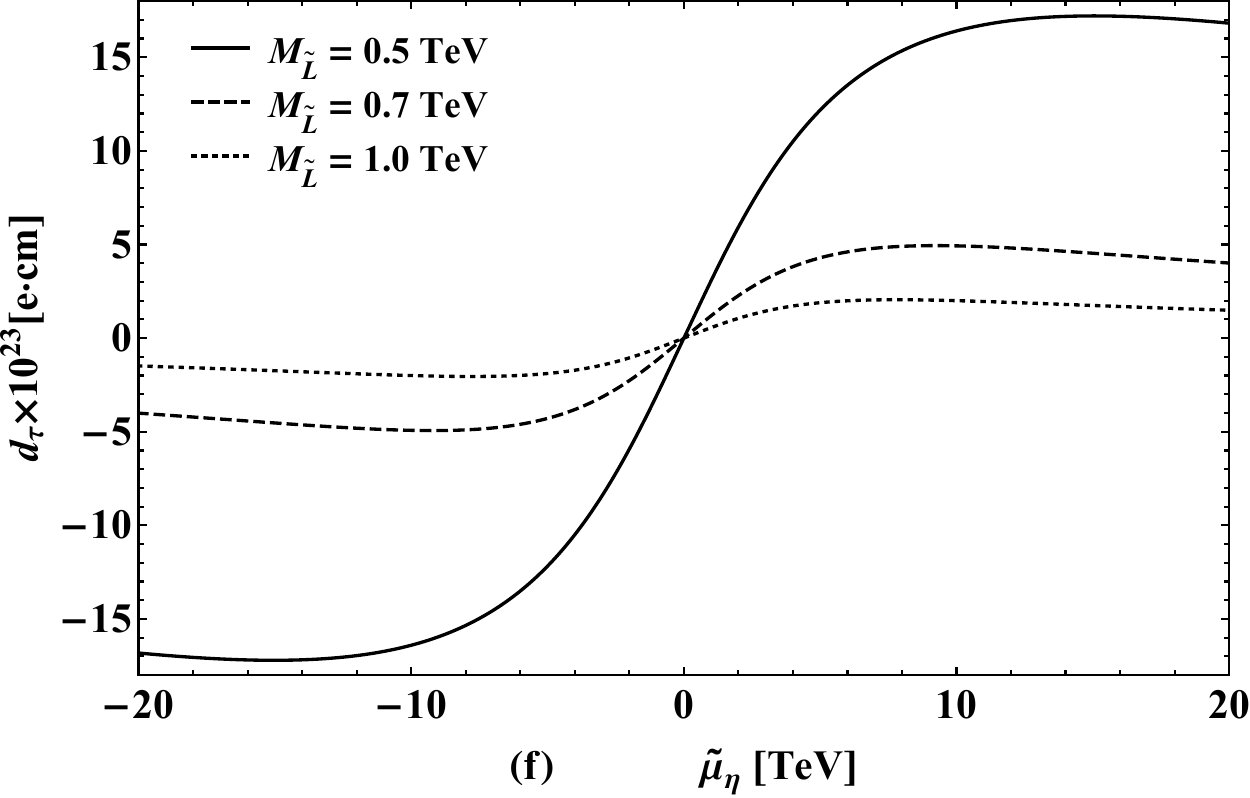}
\vspace{0cm}
\caption[]{Effects of B-LSSM specific CPV parameters on $d_\mu$, $d_\tau$. The panels display the dependence of $d_\mu$ and $d_\tau$ on $\tilde M_{BB'}$ (a, b), $\tilde M_{BL}$ (c, d), and $\tilde \mu_\eta$ (e, f), with the respective CPV phases fixed at $\theta_{BB'} = 0.5\pi,\; \theta_{BL} = 0.5\pi,\; \theta_{\mu_\eta} = 0.5\pi$. The solid, dashed, and dotted lines represent the slepton mass parameters $M_{\tilde{L}} = 0.5, 0.7, 1.0\;\text{TeV}$, respectively.}
\label{Fig4}
\end{figure}

As mentioned above, the B-LSSM introduces three specific CPV parameters: $M_{BB'},\;M_{BL}$ and $\mu_\eta$. To see their contributions to $d_\mu,\;d_\tau$ numerically, we take $\theta_{BB'}=0.5\pi$ and plot $d_\mu$ and $d_\tau$ as functions of $\tilde M_{BB'}$ in Figs.~\ref{Fig4}(a) and ~\ref{Fig4}(b), respectively. The solid, dashed, and dotted lines denote the results of $M_{\tilde L}=0.5,\;0.7,\;1.0\;\text{TeV}$, respectively. Similarly, $d_\mu,\; d_\tau$ versus $\tilde M_{BL}$ with $\theta_{BL}=0.5\pi$ and $\tilde\mu_\eta$ with $\theta_{\mu_\eta}=0.5\pi$ are also plotted. Here, we define $M_{BB'}\equiv \tilde M_{BB'} e^{i\theta_{BB'}}$, $M_{BL}\equiv \tilde M_{BL} e^{i\theta_{BL}}$, $\mu_\eta\equiv \tilde\mu_\eta e^{i\theta_{\mu_\eta}}$.

As can be seen from the picture that, the contributions from these B-LSSM specific CPV parameters are suppressed by heavy sleptons, because they contribute to $d_\mu$ and $d_\tau$ via the slepton-neutralino loop in Fig.~\ref{Fig1} (3). Furthermore, their contributions also exhibit a strong flavor dependence, which is evident when comparing the results in the left column (Fig.~\ref{Fig4} (a), (c), (e) for $d_\mu$) with those in the right column (Fig.~\ref{Fig4} (b), (d), (f) for $d_\tau$). Similar to the reason mentioned above, the EDM operator $\bar{\psi}\sigma_{\mu\nu}\gamma_5\psi F^{\mu\nu}$ inherently dictates a chirality flip, its radiative corrections are typically proportional to the Yukawa coupling of the external fermion. Consequently, the theoretical expectation $d_\tau / d_\mu \sim m_\tau / m_\mu$ is excellently corroborated by the numerical results.

In addition, Fig.~\ref{Fig4} shows that the effects of $\tilde M_{BB'}$, $\tilde M_{BL}$, and $\tilde \mu_\eta$ on $d_\mu$, $d_\tau$ exhibit entirely different asymptotic behaviors. The variation of the EDMs with respect to $\tilde M_{BB'}$ is highly non-monotonic, with distinct extrema emerging around $|\tilde M_{BB'}| \approx 2$ and $3$ TeV. $M_{BB'}$ describes the mixing effects between the superpartners of the gauge bosons corresponding to the two $U(1)$ gauge groups, it affects the numerical results in a much more complex manner than $M_{BL}$ and $\mu_\eta$. Figs.~\ref{Fig4} (c-f) show that $d_\mu$, $d_\tau$ initially increase as $|\tilde M_{BL}|$ and $|\tilde \mu_\eta|$ increase, reaching their respective  maximum around $|\tilde M_{BL}|\approx8$ TeV and $|\tilde \mu_\eta|\approx14$ TeV. This behavior arises from an inherent physical interplay: while the CPV effects are initially enhanced by the larger parameter magnitudes, an excessively large mass scale ($|\tilde{M}_{BL}|$ or $|\tilde{\mu}_\eta|$) yields heavy neutralinos, which ultimately suppress the total contributions in accordance with the decoupling theorem. This decoupling-induced suppression at large mass scales is also responsible for the asymptotic decay observed at large $|\tilde{M}_{BB'}|$ in Figs.~\ref{Fig4}(a) and 4(b).

\subsection{The combined predictions}

\begin{figure}
\setlength{\unitlength}{1mm}
\centering
\includegraphics[width=3.1in]{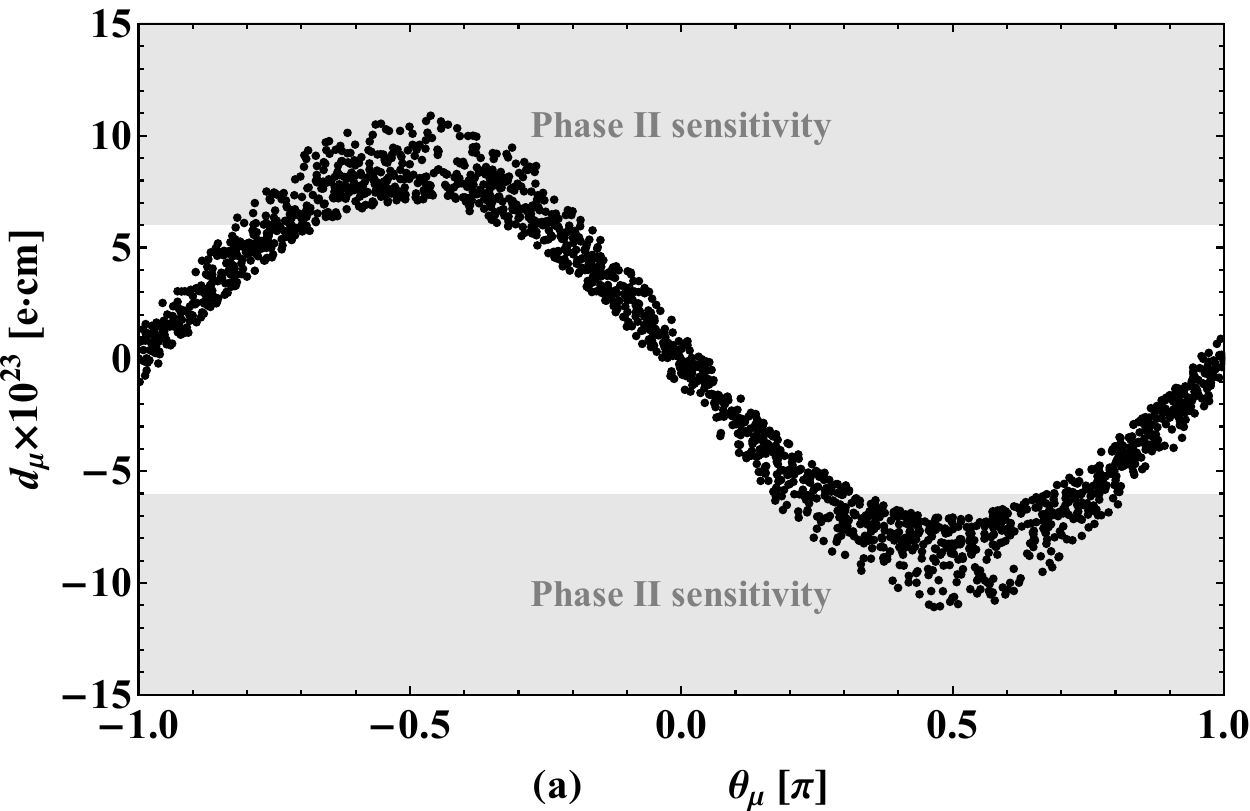}%
\vspace{0.5cm}
\includegraphics[width=3.1in]{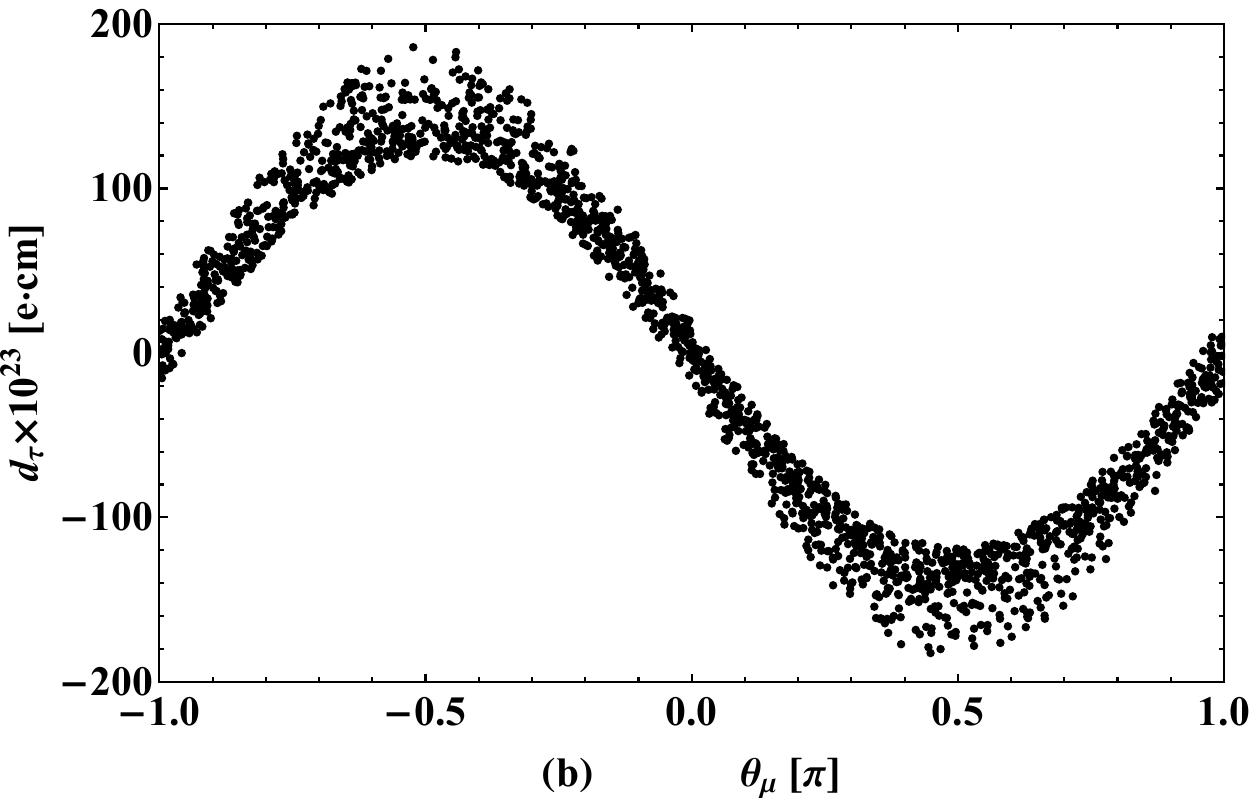}
\vspace{0cm}
\caption[]{Scatter plots of $d_\mu$ (a) and $d_\tau$ (b) versus the CPV phase $\theta_\mu$ in the B-LSSM, incorporating the combined effects of all CPV phases. The fixed mass parameters are $|\mu|=0.3\;\text{TeV}$, $\tilde M_{BB'}=3\;\text{TeV}$, $\tilde M_{BL}=8\;\text{TeV}$, and $\tilde \mu_\eta=10\;\text{TeV}$. The points correspond to a random parameter scan over $\theta_\mu, \theta_{A_l}, \theta_{BB'}, \theta_{BL}, \theta_{\mu_\eta} \in [-\pi, \pi]$. The gray shaded area denotes the sensitivity reach of the phase II experiment.}
\label{Fig5}
\end{figure}

In this subsection, we present the theoretical predictions for $d_\mu$ and $d_\tau$ by considering the combined effects of all possible CPV parameters. To illustrate regions of the parameter space with substantial EDM predictions, we apopt the benchmark values $|\mu|=0.3\;\text{TeV}$, $\tilde M_{BB'}=3\;\text{TeV}$, $\tilde M_{BL}=8\;\text{TeV}$, and $\tilde \mu_\eta=10\;\text{TeV}$ based on the preceding analysis. We then perform a random scan over the following parameter space:
\begin{eqnarray}
&&\theta_\mu\in[-\pi,\pi],\;\theta_{A_l}\in[-\pi,\pi],\;\theta_{BB'}\in[-\pi,\pi],\;\theta_{BL}\in[-\pi,\pi],\;\theta_{\mu_\eta}\in[-\pi,\pi],
\end{eqnarray}
and plot $d_\mu$ and $d_\tau$ versus $\theta_\mu$ in Figs.~\ref{Fig5} (a) and Fig.~\ref{Fig5} (b), respectively.

As clearly observed in Fig.~\ref{Fig5} (a) and Fig.~\ref{Fig5} (b), the lepton EDMs $d_\mu$ and $d_\tau$ exhibit a strong, nearly sinusoidal oscillatory dependence on the phase $\theta_\mu$. Both reach their peak magnitudes at $\theta_\mu \approx \pm 0.5\pi$, whereas the corresponding EDM predictions approach zero in the vicinity of $\theta_\mu \approx 0$ and $\pm\pi$. This behavior indicates that, within the selected parameter space, the CPV phase $\theta_\mu$ of the Higgsino mass parameter $\mu$ provides the dominant contributions to $d_\mu$ and $d_\tau$. Furthermore, the scatter points do not follow a single uniform curve but instead form a ``band'' with a specific vertical width. The band widths for $d_\mu$ in Fig.~\ref{Fig5} (a) and $d_\tau$ in Fig.~\ref{Fig5} (b) are approximately $\Delta d_\mu \sim 3 \times 10^{-23}$ e$\cdot$cm and $\Delta d_\tau \sim 50 \times 10^{-23}$ e$\cdot$cm, respectively.  Since these scatter points mainly results from the B-LSSM specific CPV parameters, it indicates they can also affect the theoretical predictions on $d_\mu$, $d_\tau$ significantly, although $\theta_\mu$ affects $d_\mu$, $d_\tau$ dominantly.

Numerically, the model yields a maximum prediction of $|d_\mu| \sim 1.0 \times 10^{-22}$ e$\cdot$cm, while the maximum $|d_\tau|$ is substantially amplified to $\sim 1.8 \times 10^{-21}$ e$\cdot$cm. The gray shaded area highlighted in Fig.~\ref{Fig5} (a) indicates the anticipated reach of the Phase II experiment for the muon EDM ($|d_\mu| \gtrsim 6 \times 10^{-23}$ e$\cdot$cm). As shown by the numerical results, the majority of the theoretical scatter points land inside this accessible region for large CPV phases within $0.25\pi \lesssim |\theta_\mu|  \lesssim 0.75\pi$. This leads to a highly significant phenomenological conclusion: the Phase II experiment will be sensitive enough to either verify or rule out regions of the B-LSSM parameter space featuring substantial $\theta_\mu$ phases. Conversely, a continued null result from future experiments would tightly restrict $\theta_\mu$ to a narrow interval around zero, unless severe fine-tuning via multi-parameter destructive interference is invoked to reconcile the model with the experimental data.

\section{Summary\label{sec4}}
In this work, we present a phenomenological study on the EDMs of the muon ($d_\mu$) and tau ($d_\tau$) within the framework of the B-LSSM. We incorporate both the general SUSY CPV sources ($\mu$, $A_l$) and the B-LSSM specific CPV parameters ($M_{BB'}$, $M_{BL}$, $\mu_\eta$), systematically analyzing their loop-induced contributions to $d_\mu$, $d_\tau$. We found that while the decoupling theorem is strictly and immediately manifested for heavy sleptons and Higgsinos, the B-LSSM specific parameters $M_{BB'}$, $M_{BL}$, $\mu_\eta$ exhibit more complex behaviors. For these specific parameters, the decoupling suppression only begins to dominate the numerical results when their respective mass scales become exceedingly large (e.g., $\gtrsim$ 20 TeV).

Furthermore, our numerical scans reveal that the theoretical predictions for $d_\mu$ and $d_\tau$ exhibit a pronounced sinusoidal dependence on $\theta_\mu$, indicating that the CPV phase of the Higgsino mass parameter $\mu$ provides the dominant contributions to $d_\mu$ and $d_\tau$. Although the B-LSSM specific CPV parameters provide subleading corrections, they can still significantly modulate the overall theoretical predictions. The B-LSSM naturally yields a muon EDM on the order of $\mathcal{O}(10^{-22})~e\cdot\text{cm}$, which substantially overlaps with the projected detection sensitivities of the recently proposed Phase II experiments. While $|d_\tau|$ can reach about $10^{-21}e\cdot\text{cm}$, such values remain challenging to observe in the near future.

\begin{acknowledgments}
The work has been supported by the National Natural Science Foundation of China (NNSFC) with Grants No. 12075074, No. 12235008, the Hebei Natural Science Foundation for Distinguished Young Scholars with Grant No. A2023201041 and the youth top-notch talent support program of the Hebei Province.
\end{acknowledgments}

\appendix

\section{Constants $C^{L(R)}_{XYZ}$ appear in our calculation \label{A1}}

The used coupling constants in the computations are
\begin{eqnarray}
&&C^L_{\bar{l}\chi^{\pm}_{i}\nu_{{\rm odd},j}}=-\frac{1}{\sqrt{2}}U_{i2}^* \sum_{b=1}^3 Z_{jb}^{I,*} \sum_{a=1}^3 U_{R,la}^{e,*} Y_{e,ab}
\end{eqnarray}

\begin{eqnarray}
&&C_{\bar{l} \chi^\pm_i \nu_{{\rm odd}, j}}^R =\frac{1}{\sqrt{2}} \left( g_2 \sum_{a=1}^3 Z_{ja}^{I,*} U_{L,la}^e V_{i1} - \sum_{b=1}^3 \sum_{a=1}^3 Y_{\nu,ab}^* Z_{j,3+a}^{I,*} U_{L,lb}^e V_{i2} \right)\\
&&C_{\bar{l} \chi^\pm_i \nu_{{\rm even}, j}}^L =  \frac{1}{\sqrt{2}} U_{i2}^* \sum_{b=1}^3 Z_{jb}^{R,*} \sum_{a=1}^3 U_{R,la}^{e,*} Y_{e,ab} \\
&&C_{\bar{l} \chi^\pm_i \nu_{{\rm even}, j}}^R =  \frac{1}{\sqrt{2}} \left( -g_2 \sum_{a=1}^3 Z_{ja}^{R,*} U_{L,la}^e V_{i1} + \sum_{b=1}^3 \sum_{a=1}^3 Y_{\nu,ab}^* Z_{j,3+a}^{R,*} U_{L,lb}^e V_{i2} \right)\\
&&C_{h_k \bar{\chi}^\pm_i \chi^\pm_j} = -\frac{1}{\sqrt{2}} g_2 \left( U_{j1}^* V_{i2}^* Z_{k2}^H + U_{j2}^* V_{i1}^* Z_{k1}^H \right),  \quad C_{h_k \bar{l} l}^* = -\frac{1}{\sqrt{2}}  Y_{e,ll} Z_{k1}^H\\[1.5ex]
&&C_{Z\bar{l}l}^L = \frac{1}{2} \Big( -g_1  \cos\Theta'_W \sin\Theta_W + g_2 \cos\Theta_W \cos\Theta'_W + (g_{YB} + g_B) \sin\Theta'_W \Big) \\[1.5ex]
&&C_{Z\bar{l}l}^R = -\frac{1}{2} \Big( 2g_1 \cos\Theta'_W \sin\Theta_W - (2g_{YB} + g_B) \sin\Theta'_W \Big)\\
&&C_{Z'\bar{l}l}^L = \frac{1}{2} \Big( \big( g_1  \sin\Theta_W - g_2 \cos\Theta_W \big) \sin\Theta'_W + (g_{YB} + g_B) \cos\Theta'_W \Big) \\[1.5ex]
&&C_{Z'\bar{l}l}^R = \frac{1}{2} \Big( 2g_1  \sin\Theta_W \sin\Theta'_W + (2g_{YB} + g_B) \cos\Theta'_W \Big) \\
&&C_{Z\bar{\chi}^\pm_i\chi^\pm_j}^L = \frac{1}{2} \Big( 2g_2 U_{j1}^* \cos\Theta_W \cos\Theta'_W U_{i1} + U_{j2}^* \big( -g_1 \cos\Theta'_W \sin\Theta_W \nonumber\\
&&\hspace{1.9cm}+ g_2 \cos\Theta_W \cos\Theta'_W + g_{YB} \sin\Theta'_W \big) U_{i2} \Big) \\[1.5ex]
&&C_{Z\bar{\chi}^\pm_i\chi^\pm_j}^R = \frac{1}{2} \Big( 2g_2 V_{i1}^* \cos\Theta_W \cos\Theta'_W V_{j1}+ V_{i2}^* \big( -g_1 \cos\Theta'_W \sin\Theta_W \nonumber\\
&&\hspace{1.9cm}+ g_2 \cos\Theta_W \cos\Theta'_W + g_{YB} \sin\Theta'_W \big) V_{j2} \Big) \\
&&C_{Z'\bar{\chi}^\pm_i\chi^\pm_j}^L = -\frac{1}{2} \Big( 2g_2 U_{j1}^* \cos\Theta_W \sin\Theta'_W U_{i1} \nonumber \\
&&\hspace{2.0cm} - U_{j2}^* \big( (g_1 \sin\Theta_W - g_2 \cos\Theta_W) \sin\Theta'_W + g_{YB} \cos\Theta'_W \big) U_{i2} \Big) \\[1.5ex]
&&C_{Z'\bar{\chi}^\pm_i\chi^\pm_j}^R = -\frac{1}{2} \Big( 2g_2 V_{i1}^* \cos\Theta_W \sin\Theta'_W V_{j1} \nonumber \\
&&\hspace{2.0cm} - V_{i2}^* \big( (g_1 \sin\Theta_W - g_2 \cos\Theta_W) \sin\Theta'_W + g_{YB} \cos\Theta'_W \big) V_{j2} \Big) \\
&&C_{\bar{l} \chi^0_i \tilde{E}_j}^L =  \Bigg( -\frac{1}{\sqrt{2}}2g_1  N_{i1}^* \sum_{a=1}^3 Z_{j,3+a}^{E,*} U_{R,la}^{e,*} - \frac{1}{\sqrt{2}}(2g_{YB} + g_B) N_{i5}^* \sum_{a=1}^3 Z_{j,3+a}^{E,*} U_{R,la}^{e,*}\nonumber\\
&&\hspace{2.0cm} - N_{i3}^* \sum_{b=1}^3 Z_{jb}^{E,*} \sum_{a=1}^3 U_{R,la}^{e,*} Y_{e,ab} \Bigg)
\end{eqnarray}

\begin{eqnarray}
&&C_{\bar{l} \chi^0_i \tilde{E}_j}^R = \frac{1}{2} \Bigg( -2 \sum_{b=1}^3 \sum_{a=1}^3 Y_{e,ab}^* Z_{j,3+a}^{E,*} U_{L,lb}^e N_{i3} 
+ \sqrt{2} \sum_{a=1}^3 Z_{ja}^{E,*} U_{L,la}^e \Big( g_1 N_{i1} + g_2 N_{i2} \nonumber\\
&&\hspace{1.7cm}+ (g_{YB} + g_B)N_{i5} \Big) \Bigg)\\
&&C_{h_k\tilde{E}_i\tilde{E}_i} = \frac{1}{4} \Bigg( -2 \Big( \sqrt{2} \sum_{b=1}^3 Z_{ib}^{E,*} \sum_{a=1}^3 Z_{i,3+a}^E T_{e,ab} Z_{k1}^H + \sqrt{2} \sum_{b=1}^3 \sum_{a=1}^3 Z_{i,3+a}^{E,*} T_{e,ab}^* Z_{ib}^E Z_{k1}^H  \nonumber \\
&&\hspace{1.7cm} + 2v_1 \sum_{c=1}^3 Z_{i,3+c}^{E,*} \sum_{b=1}^3 \sum_{a=1}^3 Y_{e,ca} Y_{e,ba} Z_{i,3+b}^E Z_{k1}^H + 2v_1 \sum_{c=1}^3 \sum_{b=1}^3 Z_{ib}^{E,*} \sum_{a=1}^3 Y_{e,ac}^* Y_{e,ab} Z_{ic}^E Z_{k1}^H \nonumber \\
&&\hspace{1.7cm} - \sqrt{2}\mu^* \sum_{b=1}^3 Z_{ib}^{E,*} \sum_{a=1}^3 Y_{e,ab} Z_{i,3+a}^E Z_{k2}^H - \sqrt{2}\mu \sum_{b=1}^3 \sum_{a=1}^3 Y_{e,ab}^* Z_{i,3+a}^{E,*} Z_{ib}^E Z_{k2}^H\Big) \nonumber \\
&&\hspace{1.7cm} + \sum_{a=1}^3 Z_{i,3+a}^{E,*} Z_{i,3+a}^E \Big( \big(2g_1^2 + g_{YB}(2g_{YB}+g_B)\big) v_1 Z_{k1}^H \nonumber \\
&&\hspace{1.7cm} - \big(2g_1^2  + g_{YB}(2g_{YB}+g_B)\big) v_2 Z_{k2}^H + 2\big(2g_{YB} g_B + g_B^2\big) (-u_{\bar{\eta}} Z_{k4}^H + u_\eta Z_{k3}^H) \Big) \nonumber \\
&&\hspace{1.7cm} + \sum_{a=1}^3 Z_{ia}^{E,*} Z_{ia}^E \Big( -\big( - g_2^2 + g_{YB} g_B + g_1^2 + g_{YB}^2\big) v_1 Z_{k1}^H \nonumber \\
&&\hspace{1.7cm} + \big( - g_2^2 + g_{YB} g_B + g_1^2 + g_{YB}^2\big) v_2 Z_{k2}^H - 2\big( g_{YB} g_B + g_B^2\big) (-u_{\bar{\eta}} Z_{k4}^H + u_\eta Z_{k3}^H) \Big) \Bigg) \\[2ex]
&&C_{Z\tilde{E}_i\tilde{E}_i} = \frac{1}{2} \Bigg( \Big( -g_1  \cos\Theta'_W \sin\Theta_W + g_2 \cos\Theta_W \cos\Theta'_W + (g_{YB} + g_B) \sin\Theta'_W \Big) \sum_{a=1}^3 Z_{ia}^{E,*} Z_{ia}^E \nonumber \\
&&\hspace{1.7cm} + \Big( -2g_1  \cos\Theta'_W \sin\Theta_W + (2g_{YB} + g_B) \sin\Theta'_W \Big) \sum_{a=1}^3 Z_{i,3+a}^{E,*} Z_{i,3+a}^E \Bigg)\\
&&C_{Z'\tilde{E}_i\tilde{E}_i} = \frac{1}{2} \Bigg( \Big( \big( g_1  \sin\Theta_W - g_2 \cos\Theta_W \big) \sin\Theta'_W + (g_{YB} + g_B) \cos\Theta'_W \Big) \sum_{a=1}^3 Z_{ia}^{E,*} Z_{ia}^E \nonumber \\
&&\hspace{1.7cm} + \Big( 2g_1  \sin\Theta_W \sin\Theta'_W + (2g_{YB} + g_B) \cos\Theta'_W \Big) \sum_{a=1}^3 Z_{i,3+a}^{E,*} Z_{i,3+a}^E \Bigg)
\end{eqnarray}

In the above expressions, we have $Y_e=\text{diag}\sqrt2(m_e,m_\mu,m_\tau)/v_1$, $U_L^e=U_R^e=\text{diag}(1,1,1)$, and the matrices $U, V, Z^E, Z^I, Z^R$, and $N$ are the rotation matrices that relate the gauge eigenstates to the mass eigenstates. Their explicit definitions are as follows:

\begin{eqnarray}
&&U^* m_{\tilde{\chi}^-} V^{\dagger}=m^{dia}_{\tilde{\chi}^-},\\
&&Z^E m^2_{\tilde{l}} Z^{E,\dagger}=m^{dia}_{2,\tilde{l}},\\
&&Z^I m_{\tilde \nu_{\rm odd}}^2 Z^{I,\dagger}=m^{dia}_{2,m_{\tilde \nu_{\rm odd}}},\\
&&Z^R m_{\tilde \nu_{\rm even}}^2 Z^{R,\dagger}=m^{dia}_{2,m_{\tilde \nu_{\rm even}}},\\
&&N^* m_{\tilde{\chi}^0} N^{\dagger}=m^{dia}_{\tilde{\chi}^0},
\end{eqnarray}
where $m_{\tilde{\chi}^-}$ is the mass matrix of charginos.

\bibliography{refs}

\end{document}